\documentclass[a4paper,12pt]{article}
\usepackage[utf8]{inputenc}
\usepackage{cancel}
\usepackage{tikz}
\usepackage{ulem}
\usepackage{amsfonts}
\usepackage{amssymb}
\usepackage{graphicx}
\usepackage{amsmath}
\usepackage{enumerate}
\usepackage{mathtools}
\usepackage{subfig}
\usepackage{color}
\usepackage{tikz}
\usepackage{float}
\usepackage{here}
\usepackage{cite}
\usepackage{mathrsfs}
\usepackage{float,epsfig}
\usepackage{dcolumn}
\usepackage{graphicx}
\usepackage{bm}
\usepackage{amsmath,amssymb,amsthm}
\usepackage[colorlinks=true,linkcolor=blue,citecolor=red]{hyperref}
\usepackage{multirow}
\usepackage[toc,page]{appendix}
\usetikzlibrary{arrows.meta}
\usetikzlibrary{bending}
\usetikzlibrary{calc}
\newcommand{\be}{\begin{equation}}
\newcommand{\ee}{\end{equation}}
\newcommand{\bea}{\setlength\arraycolsep{2pt} \begin{eqnarray}}
\newcommand{\eea}{\end{eqnarray}}

\def\0{{\sst{(0)}}}
\def\1{{\sst{(1)}}}
\def\2{{\sst{(2)}}}
\def\3{{\sst{(3)}}}
\def\4{{\sst{(4)}}}
\def\5{{\sst{(5)}}}
\def\6{{\sst{(6)}}}
\def\7{{\sst{(7)}}}
\def\8{{\sst{(8)}}}
\def\sst#1{{\scriptscriptstyle #1}}

\makeatletter \@addtoreset{equation}{section}

\definecolor{lime}{HTML}{A6CE39}

\begin{document}

\title{{\normalsize \textbf{\Large   DFT   Studies  of 2D Materials Inspired by  Lie Algebras }}}
\author{ {\small Adil Belhaj$^1$\footnote{a-belhaj@um5r.ac.ma}, Salah Eddine Ennadifi$^2$\footnote{ennadifis@gmail.com}\thanks{%
Authors in alphabetical order.} \hspace*{-8pt}} \\
{\small $^1$ D\'{e}partement de Physique, \'Equipe des Sciences de la
mati\`ere et du rayonnement, ESMaR}\\
{\small Facult\'e des Sciences, Universit\'e Mohammed V de Rabat, Rabat,
Morocco} \\
{\small $^2$ LHEP-MS, Facult\'e des Sciences, Universit\'e Mohammed V de
Rabat, Rabat, Morocco } }
\maketitle

\begin{abstract}
Inspired by the root systems of Lie algebras of rank 2, we propose a
mathematical method to engineer new 2D materials with double periodic
structures tessellating the plane. Concretely, we investigate two geometries relaying on  squares and hexagons exhibiting  the $D_4 \times D_4$ and $D_6
\times D_6$ dihedral group invariances, respectively. Due to lack of
empirical verifications of such double configurations, we provide a
numerical investigation by help of the open source quantum espresso.
Motivated by hybrid structures of  the graphene,  the silicene, and the  germanene, we
investigate two models involving   the $D_4 \times D_4$ and $D_6 \times D_6$
dihedral symmetries which we refer to as Si4Ge4 and Si6C6 compounds,
respectively. For simplicities, we study only the opto-electronic physical
properties  by applying an electromagnetic source propagating in linear and
isotropic mediums.    Among others,  we find  that such  2D materials  exhibit metallic
behaviors with certain optical features.  Precisely,  we  compute  and discuss  the  relavant  optical quantities including  the dielectric function,
the absorption spectra,  the refractive index,  and   the reflectivity. We believe that the Lie algebra inspiration of such 2D
material studies, via density functional theory techniques, could open new
roads to think about higher dimensional cases.
\end{abstract}
 \newpage

\newpage

%



\section{Introduction}

High Energy Physics (HEP) concerns the study of the Universe aspects at the
most fundamental level via empirical and rational investigation methods
including simulations using developed numerical codes \cite{1}. The Standard
Model of particle physics (SM) and beyond aim to describe  the fundamental
laws of the nature by approaching the relevant elementary components of
matter and  interactions which govern them  by help of a beautiful
concept called symmetry \cite{S1,S2,S3,S4}. The latter offers explanations
for many questions  on  physical properties of matter, and can 
shed  some lights on other veiled phenomena of the nature ranging from the 
deep-inside of matter at the level of quarks and gluons to large scales at
the level of big structures of the Universe \cite{S5,S6,S7}. To  study
such a primordial concept, Lie algebras and group theory are considered as
the crucial mathematical frameworks \cite{3,3aa,3a,3a1,3a2,3b,3c}. These
theories have been  exploited in many areas, comprising  Solid State Physics (SSP)
dealing with various behaviors of lower dimensional materials \cite{3e,3f}.
For instance, an experimental evidence for the exceptional $E_{8}$ Lie
algebra has been provided in the study of the ferromagnetic Ising-chain CoNb$%
_{2}$O$_{6}$ materials \cite{4,4a}. This charming Lie algebra has appeared
naturally in higher dimensional supergravity theories such as superstrings
and M-theory via compatification scenarios generating models in four
dimensional spacetime geometries  \cite{5ab,5ac,5,5a,5b,5c,5d,5e}.

In connection with the development of new materials, several studies have
been elaborated in order to investigate the physical properties of two
dimensional (2D) materials \cite{6,6a1,6a2,6a3,6a4,6a5,6a6,6a7,6a8,6a9,6a10,6a11}. The extensively dealt with one is
graphene, using a variety of computational methods with suitable
approximations. This material is a single-layer carbon crystal with 
the hexagonal geometry \cite{7}. The energy spectrum of this material has a
special structure in which the valence and the conduction bands intersect at
the Dirac points to produce a semi-metal \cite{8,8a,8b,8c}. Low-energy
excitations are described by a pair of two-component massless fermions or,
equivalently, by four-component fermion wave functions verifying the Dirac
equation describing massless particles. Various properties of 2D materials
have been investigated using the Density Functional Theory (DFT) being
considered as one of the most exploited computational methods based on the 
theoretical foundations of quantum mechanics \cite{9}. Many approximations
have been  explored to get results which could be compared  with 
material empirical findings. Concretely, the Local Density Approximation
(LDA) has been used  in many 2D material activities \cite{10}. This
approximation  consists in assuming that the exchange and the
correlation potential is a function of the local electron density. In such
an approximation, however, the exchange and the correlation potential depend
only on the local density and not on the local variations of this density.
Moreover, it has been remarked that it is well adopted only to homogeneous
backgrounds. To go beyond such particular situations, other approximations
have been emerged in DFT computational methods. For instance, the
Generalized Gradient Approximation (GGA) has been explored to improve the
accuracy on the value of the total energy including other energetic
contributions \cite{11}.

Recently, it has been remarked that the connection  between  SSP and HEP
has been shown to be a beautiful investigation direction. The graphene
results have opened interesting gates for elaborating dual and interplayed
scenarios. Albeit the link concerns these two extremely different domains,
the corresponding interplay has provided a road for the non trivial
solutions of the Einstein equation to be relevant pieces in the analogue
gravity \cite{12}. In connection with  the DFT calculations by use of  the Quantum
Espresso Code (QEC), the Black Hole (BH) physics has been exploited to study
lower dimensional materials \cite{13}.


The curiosity on such connected subjects could trigger  other   investigation
issues. To provide other links, indeed, many open questions could be
addressed inspired by the HEP paradigm. Looking for alternative grounds
between HEP and  SSP, the hadronic structure where the
hexagonal one appearing in the quark particle building models from the root
system of the $su(3)$ Lie algebra via  appropriate  representations including the
adjoint one could be placed at the center  of the inspiration. Indeed, the
gatherings of quarks in the SM mesonic $\mathbf{q}\overline{\mathbf{q}}$ and the 
baryonic $\mathbf{qqq}$ states appear to respect a certain geometry
belonging to the $SU(2)\times SU(3)$ Lie group of the weak and  the strong gauge
symmetry, respectively\cite{S8}. Extending the quark combinations with
respect to such symmetries as well as to other possible underlying
symmetries leads to a general quark structure giving rise to new hadronic
states by means of polygon geometries according to the corresponding Lie
algebras \cite{S6}. Supported by such a bridging scenario, it could be possible to use
known results of lower rank Lie algebras to supply new periodic geometries
via the root systems. 

The present study concerns the following investigation
direction 
\begin{equation*}
\mbox{Lie algebras}\rightarrow \mbox{2D materials}.
\end{equation*}%
At first sight, this route seems strange since one could identify symmetries
from  the atom organizations in materials. However, here, we follow the opposite
direction. Inspired by the root systems of the Lie algebras with rank 2, we
propose a mathematical method to engineer new 2D materials with double
periodic structures tessellating the plane. Concretely,  we  first model two geometries relaying on squares and hexagons involving the  $
D_{4}\times D_{4}$ and $D_{6}\times D_{6}$ dihedral group invariances,
respectively.  We refer to such materials  as Si4Ge4 and Si6C6 compounds, respectively.  Due to lack of empirical verifications of such double
structures, we perform  a numerical investigation by help of the open source
QEC with GGA approximations. We expect that the present work could be
extended to other DFT numerical codes which could   provide  either refined findings  or at
least similar ones.   For simplicity reasons, we present only the
opto-electronic physical behaviors by applying an electromagnetic source
propagating in linear isotropic mediums.   Among others,   the obtained   2D materials manifest   metallic
behaviors with certain optical features.   Concretely,  we   calculate  and examine   the  relavant  optical quantities including the dielectric function, 
the absorption spectra,  the refractive index,  and   the  reflectivity.  In addition to these findings, we
anticipate that the present Lie algebra inspired models could open new roads
to approach higher dimensional materials.

The organization of this work is as follows. In section 2, we give a concise
review on the root systems of the rank 2 Lie algebras. In section 3, we
bridge the Lie algebras to 2D materials via a specific correspondence. In
sections 4 and 5, we discuss certain  physical properties of 2D materials with periodic
double structures using DFT  simulation calculations. The last section is devoted
to conclusions  and certain open questions.

\section{ Root system geometries of rank two Lie algebras}

In physics, the geometry of root systems has been considered as a relevant
ingredient. Specifically, it provides the crucial role played by Lie
algebras in  SM and higher dimensional physical models, including string
theory and related dual theories \cite{18,19}. In this way, the root systems
have been explored to partially solve many physical problems. Particular
emphasis has been put on the hexagonal geometry appearing naturally in
particle representations of the SM considered as a good candidate to deal
with  the fundamental interactions of the Universe in an elegant way. It has been
observed that the hexagonal structure also appears in lower dimensional
theories, including SSP  used to treat graphene-like models.
These structures share similarities with the hexagonal root systems of Lie
symmetries. Before arriving to what we are looking for, it is recalled that
a Lie symmetry $L$ is a vector space provided with an antisymmetric bilinear
bracket $[,]$ : $L\times L\rightarrow L$. This mapping  operation should satisfy some
conditions including the Jacobi identity ($([x,[y,z]]+[z,[x,y]]+[y,[z,x]]=0)$%
) which replaces the associativity property \cite{3,3aa,3c}.    Certain details  on Lie algebras have been added in the appendix section.  A close
inspection shows that any finite Lie algebra has a special sub-algebra called
Cartan sub-algebra usually denoted by $H$. The latter is the maximal abelian
Lie sub-algebra, being relevant in the representation theory of Lie
algebras. By help of such an abelian sub-algebra, $L$ can be written as the
direct sum of $H$ and a vector sub-space $E_{\Delta }$. In this way, one
writes 
\begin{equation}
L=H\oplus E_{\Delta }.
\end{equation}%
Moreover, $E_{\Delta }$ can be factorized in terms of one-dimensional vector
spaces $L_{\alpha }$ given by 
\begin{equation}
E_{\Delta }=\oplus _{\alpha }L_{\alpha }.
\end{equation}%
For $x$ inside $L$ and $\alpha $ ranges over all elements of the dual of $H$%
, $L_{\alpha }$ is defined by 
\begin{equation}
L_{\alpha }=\{x\in L|[h_a,x_j]=\alpha_{aj} (x_j)x_j\}
\end{equation}%
where $h_a$ are generators of $H$.  In  the Lie algebra terminology,  $\alpha_{aj}=\alpha$ are called roots belonging to the
associated root system $\Delta $. The latter is a subset of an euclidean
space $E$ verifying specific conditions. Indeed, $\Delta $ does not contain
a null vector and involves a finite set generating $E$. If $\alpha $ is a
root, only $-\alpha $ is a root. For any root $\alpha $, one defines a
reflection $\sigma _{\alpha }$, where one has $\sigma _{\alpha }(\beta
)=\beta -\frac{\beta .\alpha }{\alpha ^{2}}\alpha $ which leaves $\Delta $
invariant. It turns out that the scalar quantity $\frac{\beta .\alpha }{%
\alpha ^{2}}$ should be an integer. An examination reveals that $\Delta $
carries crucial information which could be explored to bridge the associated
structure to physics. For later use, a special focus  can be put on rank 2
Lie algebras having only two simple roots $\alpha _{1}$ and $\alpha _{2}$
being separated by an angle $\theta _{12}$. For finite Lie algebras, it is
denoted that the rank is exactly the dimension of the Cartan sub-algebra $H$. For instance, the rank 2 Lie algebras are known by the following
constraint 
\begin{equation}
\dim H=\dim L-|\Delta |=rank\;L=2
\end{equation}%
where $|\Delta |$ is the number of the roots defining  the algebraic
structure of  the Lie algebra $L$.  A nearby examination reveals that
there is a nice classification of the rank 2 Lie algebras which leads to
four solutions with planar root systems. The classification is illustrated
in the table (\ref{tab41}). 
\begin{table}[th]
\begin{center}
\begin{tabular}{|c|c|c|c|c|}
\hline
Lie algebra & rank & $\theta_{12}$ & $|\bigtriangleup| $ & Geometry of $%
\Delta $ \\ \hline
$su(2)\oplus su(2) $ & 2 & $90^{\circ}$ & 4 & single square \\ \hline
$su(3) $ & 2 & $120^{\circ}$ & 6 & single hexagon \\ \hline
so(5) & 2 & $135^{\circ}$ & 8 & double square \\ \hline
G2 & 2 & $150^{\circ}$ & 12 & double hexagon \\ \hline
\end{tabular}%
\end{center}
\caption{Classification of rank two Lie algebras.}
\label{tab41}
\end{table}
A close inspection shows that this classification could provide periodic
polygons tessellating the full plane generating supercell structures in 2D
material physics. This leads to a bridge between the root system $\Delta $
of rank 2 Lie algebras and the atom organizations of 2D materials.  

A close examination on such Lie algebras shows  that one has two Lie symmetries being $ so(5)$  and  $G2$ ones involving double structures in the associated root systems. These geometries will be exploited to  engineer new 2D materials  going beyond the usual one involving the single periodic structure  of  the graphe-like models.  For square  geometries, we will establish a corresponding between  the  $so(5)$ Lie algebra and the atom configurations  in the  corresponding 2D materials.   Inspired by  the $G2$ Lie algebra, similar arguments will be used to engineer  the double  hexagonal   structure for 2D materials.    Before going ahead, we first give an illustration  of  such root systems  with double  structures.

\subsection{$so(5)$ root system}
\begin{figure}[th]
\begin{center}
\begin{tikzpicture}[
    -{Straight Barb[bend,
       width=\the\dimexpr10\pgflinewidth\relax,
       length=\the\dimexpr12\pgflinewidth\relax]},
  ]
    \foreach \i in {0,1,2,3,4} {
      \draw[thick, green] (0, 0) -- (\i*90:2);
      \foreach \x in {0,90,133,270} { 
      	-- (\x-90:2)--(\x:2) 
      } -- cycle;
      \draw[thick, blue] (0, 0) -- (45 + \i*90:2.828427);
      
    }
    
  \begin{scope}[yshift=0cm,rotate=90]
    \draw[green, thick]  (0:2) --(90:2)--(180:2)--(270:2)--cycle;
  \end{scope}
    \begin{scope}[yshift=0cm,rotate=45]
    \draw[blue, thick]  (0:2.828427) --(90:2.828427)--(180:2.828427)--(270:2.828427)--cycle;
    
  \end{scope}
    \node[right,green] at (2, 0) {$\alpha_{1}$};
    \node[right,green] at (-2.7, 0) {-$\alpha_{1}$};
    \node[right,green] at (-0.81, 2.3) {($\alpha_{1}$+$\alpha_{2}$)};
    \node[right,green] at (-0.9, -2.3) {-($\alpha_{1}$+$\alpha_{2}$)};
    \node[above left, inner sep=.2em,blue] at (1*135:2.8) {$\alpha_{2}$};
    \node[above left, inner sep=.2em,blue] at (-1*130:3.1) {-($2\alpha_{1}$+$\alpha_{2}$)};
    \node[above left, inner sep=.2em,blue] at (-1*44:3.5) {-$\alpha_{2}$};
    \node[above left, inner sep=.2em,blue] at (1*30:4.1) {($2\alpha_{1}$+$\alpha_{2}$)};
    
  \end{tikzpicture}
\end{center}
\caption{Root system geometry of the $so(5) $ Lie algebra.}
\label{F1}
\end{figure}
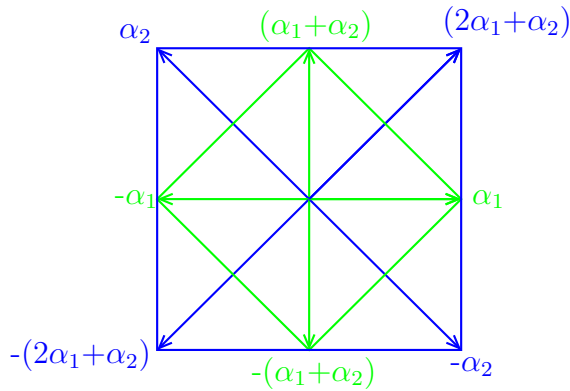

 To engineer  2D materials with double squares, we exploit   
the root system of  the $so(5)$Lie algebra.  The latter has been explored in many investigations including
BH and string theory via the AdS/CFT correspondence \cite{20}. It has been
remarked that the $so(5)$ Lie algebra is defined by two simple roots $\alpha
_{1}$ and $\alpha _{2}$, with different lengths, separated by $\theta
_{12}=135^{\circ }$. This angle can be split as follows 
\begin{equation}
135^{\circ }=90^{\circ }+45^{\circ }
\end{equation}%
where $90^{\circ }$ corresponds to the single square root system defining
the $su(2)\oplus su(2)$ Lie algebra. The associated root geometry involves
two squares with different sizes rotated by $45^{\circ }$. The small one is
generated by $\{\pm \alpha _{1},$ $\pm (\alpha _{1}+\alpha _{2})\}$.
However, the big one is formed by $\{\pm \alpha _{2},$ $\pm (2\alpha
_{1}+\alpha _{2})\}$. This double structure is illustrated in Fig.\ref{F1}. 

\subsection{$G2$ root system}
 Here, we  provide the  $G2$ root system which will be exploited  to  generate the double hexagonal structure in 2D materials.  It is recalled that  the $G2$ Lie algebra is a special symmetry playing a crucial role in the
M-theory compactification \cite{20,21}. It replaces the $su(3)$ Lie algebra in
the Calabi-Yau string compactification scheme. The associated Lie group has
been needed to provide a minimal supersymmetric model from the M-theory
dimensional reduction. Roughly, the  $G2$ Lie   algebra contains two simple roots $%
\alpha _{1}$ and $\alpha _{2}$, with different lengths, separated by $\theta
_{12}=150^{\circ }.$ As the previous model, this angle can be split as
follows 
\begin{equation}
150^{\circ }=120^{\circ }+30^{\circ }
\end{equation}%
where the angle $120^{\circ }$ is related to the single hexagon root system
defining the $su(3)$ Lie algebra. The corresponding root geometry has two
hexagons with different sizes rotated by $30^{\circ }$. The small one is
generated by $\{\pm \alpha _{1},\pm (\alpha _{1}+\alpha _{2}),\pm (2\alpha
_{1}+\alpha _{2})\}$. However, the big one is formed by $\{\pm \alpha
_{2},\pm (3\alpha _{1}+\alpha _{2}),\pm (3\alpha _{1}+2\alpha _{2})\}$. This
double hexagonal structure is illustrated in Fig.\ref{F2}.

\begin{figure}[!ht]
\begin{center}
\begin{tikzpicture}[
    -{Straight Barb[bend,
       width=\the\dimexpr10\pgflinewidth\relax,
       length=\the\dimexpr12\pgflinewidth\relax]},
  ]
    \foreach \i in {0, 1, ..., 5} {
      \draw[thick, green] (0, 0) -- (\i*60:2);
      \draw[green, thick] (0, 0) 
      \foreach \x in {60,120,...,360} { 
      	-- (\x-60:2)--(\x:2) 
      } -- cycle;
      \draw[blue, thick] (0, 0) 
      \foreach \x in {30,90,...,330} { 
      	-- (\x-60:3.464101615) --(\x:3.464101615) 
      } -- cycle;
      
      \draw[thick, blue] (0, 0) -- (30 + \i*60:3.464101615);
      
    }
    \draw[thin, black] (1, 0) arc[radius=1, start angle=0, end angle=1*30];
    \node[right,green] at (2, 0) {$\alpha_{1}$};
    \node[right,green] at (0.4, 2.1) {($2\alpha_{1}$+$\alpha_{2}$)};
    \node[right,green] at (-2, 2.1) {($\alpha_{1}$+$\alpha_{2}$)};
    \node[right,green] at (-2.7, 0) {-$\alpha_{1}$};
     \node[right,green] at (0.4, -2.1) {-($\alpha_{1}$+$\alpha_{2}$)};
     \node[right,green] at (-2, -2.1) {-($2\alpha_{1}$+$\alpha_{2}$)};

    \node[above left, inner sep=.2em,blue] at (5*30:3.5) {$\alpha_{2}$};
    \node[above left, inner sep=.2em,blue] at (2.545*30:3.7) {($3\alpha_{1}$+$2\alpha_{2}$)};
    \node[above left, inner sep=.2em,blue] at (7.1*30:3.7) {-($3\alpha_{1}$+$\alpha_{2}$)};
    \node[above left, inner sep=.2em,blue] at (9.4*30:4.2) {-($3\alpha_{1}$+$2\alpha_{2}$)};
    \node[above left, inner sep=.2em,blue] at (11.1*30:4.3) {-$\alpha_{2}$};
    \node[above left, inner sep=.2em,blue] at (12.6*30:5.1) {($3\alpha_{1}$+$\alpha_{2}$)};

    \node[right] at (15:1) {$30^{\circ}$};
    
  \end{tikzpicture}
\end{center}
\caption{Root system geometry of the $G2$  Lie algebra.}
\label{F2}
\end{figure}

\section{ 2D material modeling  from root system geometries}

Inspired by Lie algebra results, we could engineer certain 2D materials from
the root system geometries. This may offer a new road  to  investigate  such
materials by using a dictionary between the root systems of the 
Lie algebras and the atom organizations in lower dimensional materials. In
this way, the atoms are associated with non-zero roots of the involved Lie
algebras. According to the previous table, we can distinguish two kinds of
2D materials involving whether single structures or double structures
generated by two simple roots $\alpha _{1}$ and $\alpha _{2}$ separated by
the angle $\theta _{12}$. The first class corresponds to simply laced Lie
algebras required by 
\begin{equation}
||\alpha _{2}||=||\alpha _{1}||
\end{equation}%
being $su(2)\oplus su(2)$ and $su(3)$ Lie symmetries relaying on the single
square and the single hexagon structures, respectively. In this case, $%
\alpha _{1}$ and $\alpha _{2}$ generate the same polygon structure.
Concretely, the fundamental length parameters $a=b$ can be identified with
the length of the simple roots 
\begin{equation}
a=b=||\alpha _{1}||=||\alpha _{2}||.
\end{equation}%
Physically, the general material configurations with a flat geometry can be
derived by tessellating the full plane via a single structure in 2D material
physics. For  the $su(3)$ Lie algebra, the associated single structure is the hexagonal one
with the $D_{6}$ dihedral group symmetry. This engineers the ordinary
structures, appearing in graphene like-models including  the silicene and the 
germanene. The second class, being the more attractive and new one, concerns
the double structure inspired by the non-simply laced Lie algebras required
by the following constraints on the simple roots 
\begin{equation}
||\alpha _{2}||=2|\cos (\frac{180^{\circ }}{k})|||\alpha _{1}||
\label{fundamentalequation}
\end{equation}%
where $k$ takes two values 4 and 6 associated with the  $so(5)$ and  the $G{2}$ Lie
algebras, respectively. Contrary to the single structure, $\alpha _{1}$ and $%
\alpha _{2}$ belong to two different structures shifted by the angle $\frac{%
180^{\circ }}{k}$. It is observed that each simple root of such Lie algebras
generates a single periodic structure. The smallest one is defined by the
set of roots having the same length $||\alpha _{1}||$ while the second is
the biggest one which undergoes a rotation by the angle of $\frac{180^{\circ
}}{k}$. This structure is generated by the set of roots constrained by a
length identified with $||\alpha _{2}||$.

In the 2D material side, a double geometry arises with two fundamental
length parameters $a_{1}= b_{1}$ and $a_{2}=b_{2}$ linked by 
\begin{equation}
a_{2}=2|\cos (\frac{180^{\circ} }{k})|a_{1}.  \label{fundamentalequation1}
\end{equation}
Precisely, we could engineer 2D materials with double periodic structures.
In this way, the general flat structure can be obtained by using the fact
that the double structure tessellates the entire plane forming a new
crystalline supercell by mixing the $(1 \times 1) $ and $( 2|\cos (\frac{
180^{\circ} }{k})| \times 2|\cos (\frac{180^{\circ} }{k})|)$ superstructures
separated by the angle $\frac{180^{\circ} }{k}$. We could safely say that
this geometry is invariant under the $D_k \times D_k$ dihedral group which
could be reduced to one $D_k$ factor appearing in the single structure case
corresponding to the simply laced Lie algebras. \newline

\subsection{2D double square materials inspired by the $so(5)$ Lie algebra}

\begin{figure}[th]
\begin{center}
\includegraphics[width=8cm, height=3cm]{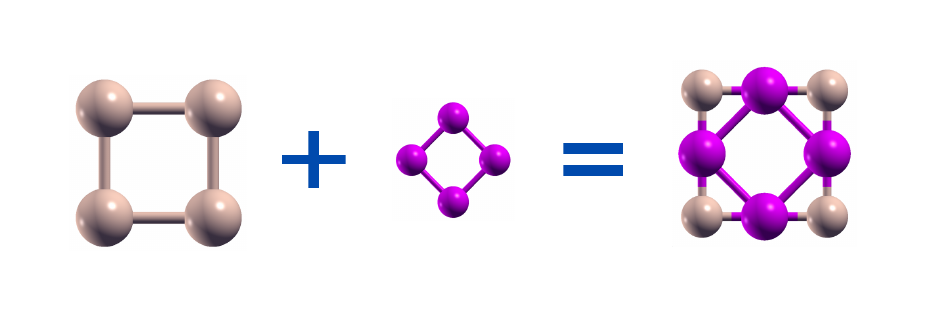}
\end{center}
\caption{Double square structure with the $D_{4}\times D_{4}$ dihedral group.}
\label{F3}
\end{figure}

\begin{figure}[h!]
\centering
\includegraphics[scale=0.5,trim=0cm 11cm 0cm
2.3cm,clip=true]{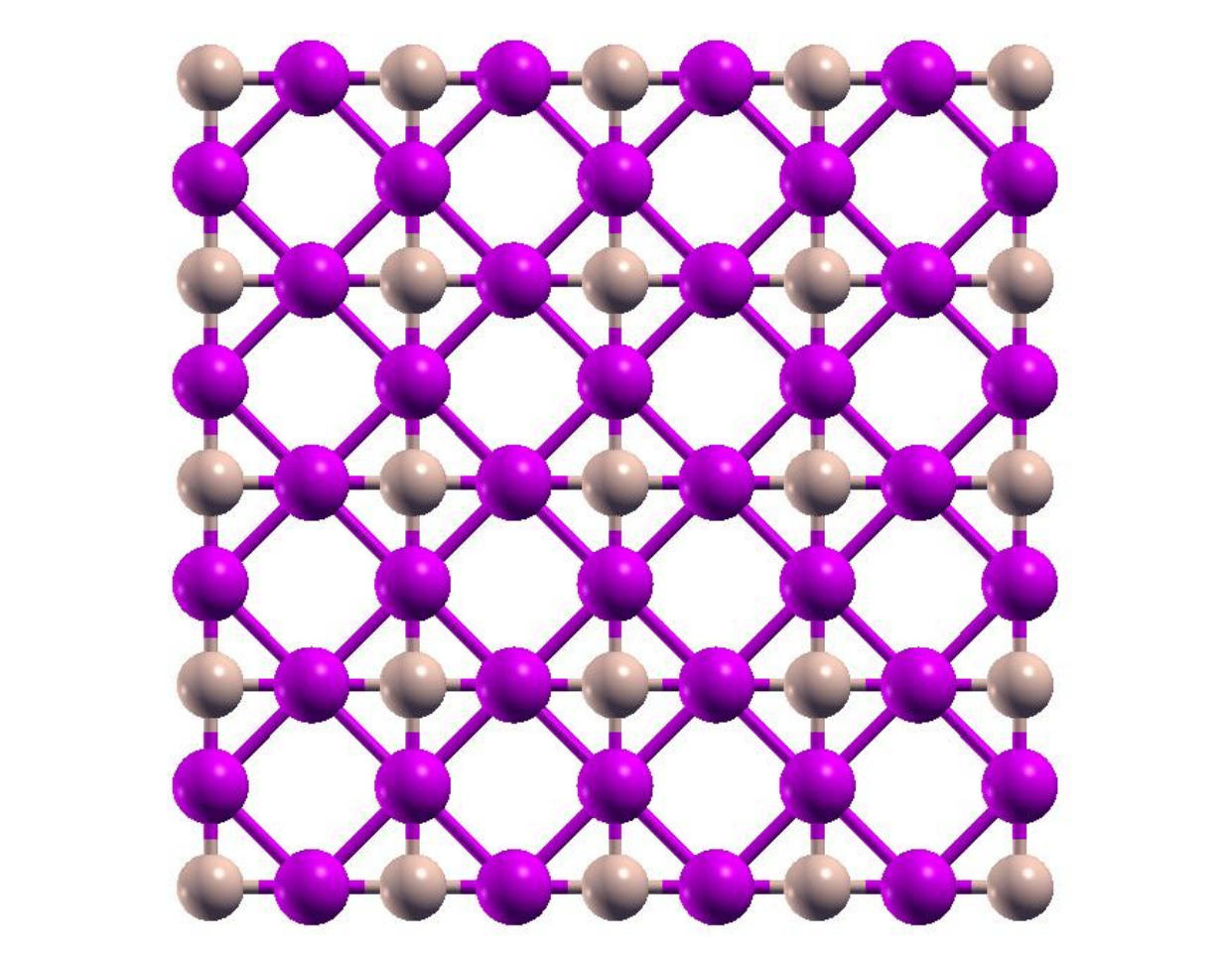}
\caption{2D double square geometry.}
\label{F4}
\end{figure}

In this subsection, we engineer  a  double square configuration inspired by
the root system of the $so(5)$ Lie algebra presented in Fig.\ref{F1}. The
fundamental geometry, involving the $D_{4}\times D_{4}$ dihedral symmetry
group, has been obtained  by combining two squares with two
different sizes rotated by $45^{\circ }$. This geometry  is illustrated in Fig.\ref{F3}%
. 
Combining two squares, we obtain a periodic  configuration  in the 2D plane as
presented in Fig.\ref{F4}.

\ 

\subsection{2D double hexagonal materials inspired by the  $G2$ Lie algebra}

\begin{figure}[th]
\begin{center}
\includegraphics[width=8cm, height=3cm]{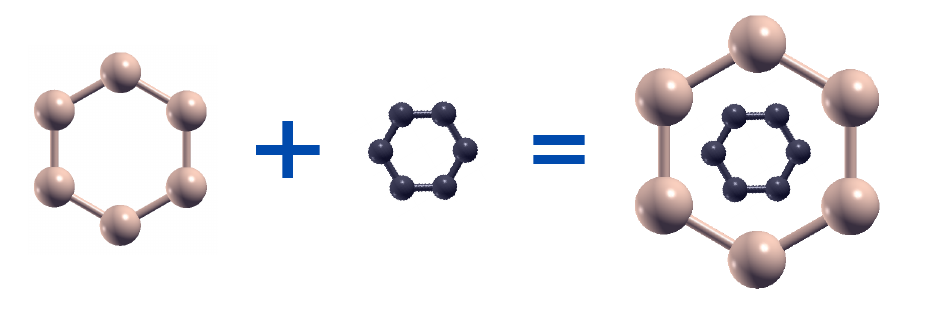}
\end{center}
\caption{Double square structure with the $D_{6}\times D_{6}$ dihedral group.
}
\label{F5}
\end{figure}

\begin{figure}[th]
\begin{center}
\includegraphics[scale=0.5,trim=0cm 13cm 0cm
2.3cm,clip=true]{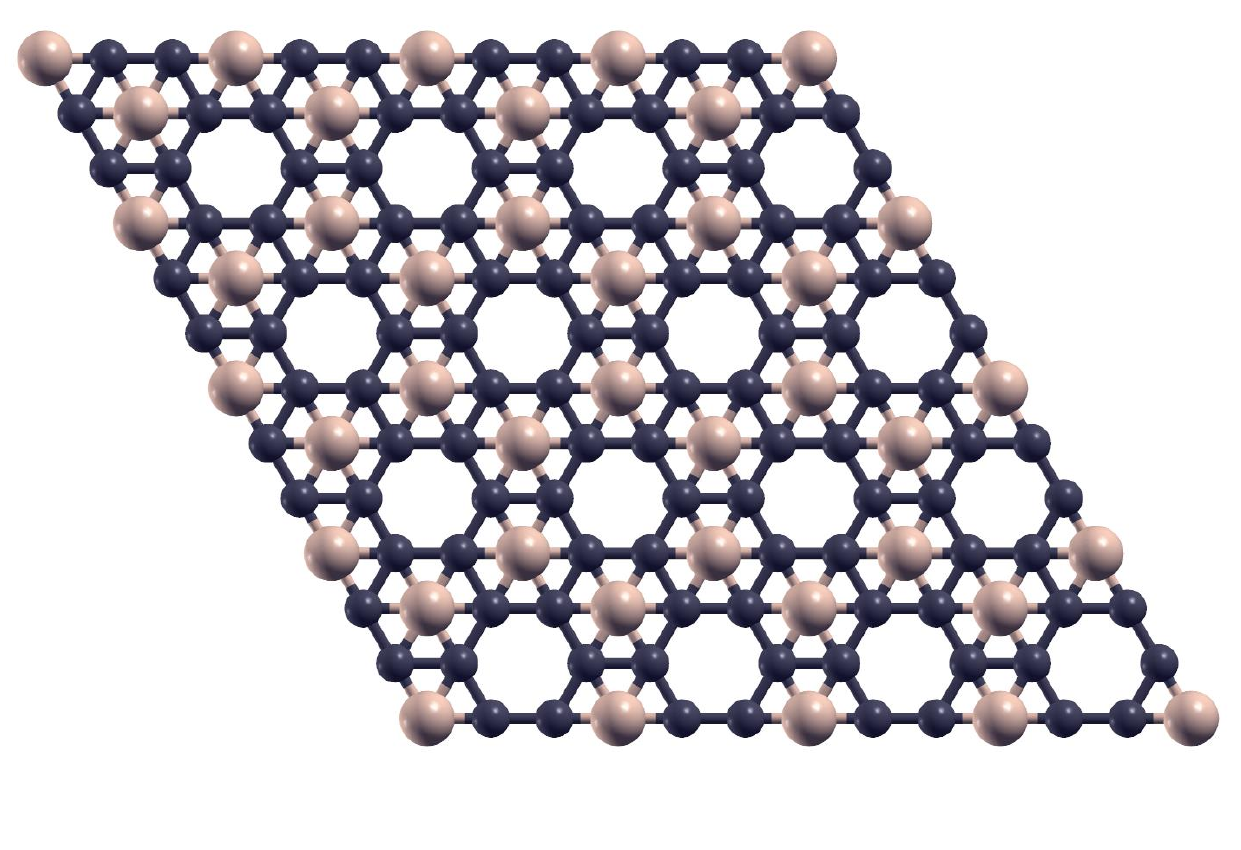}
\end{center}
\caption{2D double hexagon geometry.}
\label{F6}
\end{figure}

In this subsection, we model a double hexagonal structure inspired by the
root system of the $G2$ Lie algebra  being illustrated in Fig.\ref{F2}. The
building block geometry, having the $D_{6}\times D_{6}$ dihedral symmetry
group, is provided in Fig.\ref{F5}. 
Combining these two hexagons, we get the periodic geometry in the 2D plane
as shown in Fig.\ref{F6}. 

At this level, one may ask certain questions related to the underlying
physical properties. The  natural question concerns an empirical
test for the double structure. Honestly, we do not know the answer. However,
we could provide verifications within simulations of such 2D materials by
the help of numerical codes using DFT techniques. Concretely, we consider the
open source QEC \cite{22}.

\section{ DFT Simulations of 2D materials based on double structures}

In the remaining part of this work, we investigate certain physical behaviors of 2D materials
based on double structures using DFT computations by means of simulating codes. It
has been suggested that the ab-initio techniques have been explored to deal
with the physical properties of lower dimensional materials including  2D ones. This has been 
performed  by solving the relevant equations of quantum mechanics,
without using adjustable variables. Indeed, it has  been recalled
that DFT is a reformulation of the N-body quantum problem into a problem
which treats only  one electron density. Indeed, such methods relay only on
the equations of quantum mechanics, and not on empirical models,
observational and experimental statements. It has been based on solving the
time-independent Schr\^odinger  equation. Recently, DFT is considered
as the most widely exploited techniques for quantum calculations of certain
structures of the solids, including the electronic ones. It helps to make
accessible the computations of the ground state for a system involving
electronic multiple particles. More recently, it has been remarked that this
method has been largely explored to approach various behaviors of
interesting materials with different potential applications such as  photovoltaic and spintronic.

\begin{figure}[th]
\centering
\begin{tabular}{cc}
\centering \includegraphics[scale=0.35]{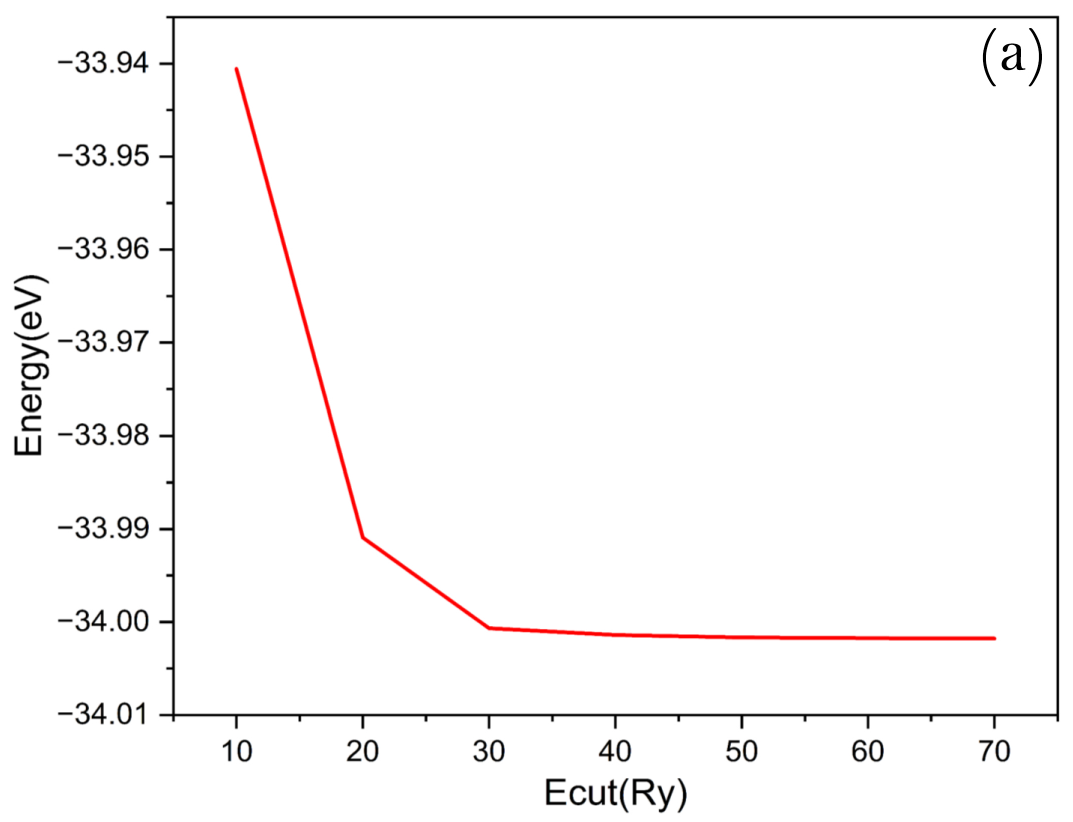} & %
\includegraphics[scale=0.35]{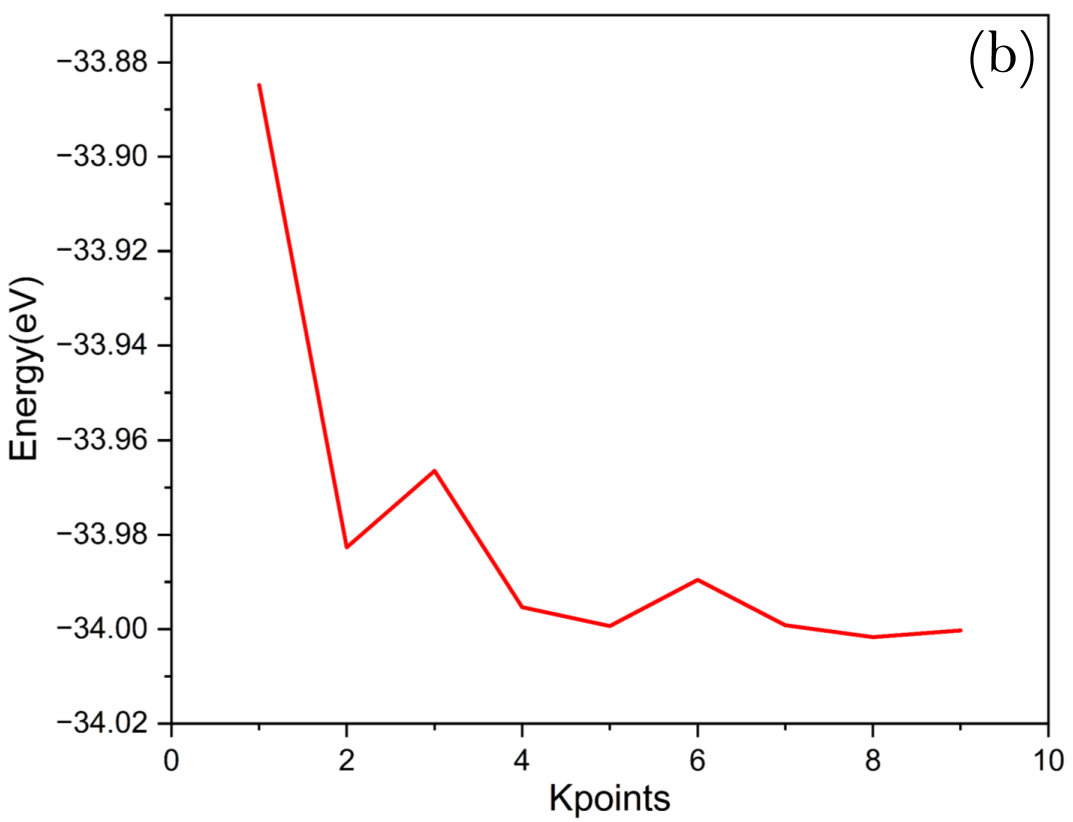}%
\end{tabular}%
\caption{(a): Total energy as  a function of cut off energy of Si4Ge4 material.
(b): Total energy as a function of K-points of Si4Ge4 material.}
\label{F7}
\end{figure}

 Roughly,  we exploit the DFT computations to support the proposed  2D materials inspired by Lie algebras.  The present  computations have been
obtained by help of QEC \cite{22}. The latter is  considered as a relevant electronic structure code.  Precisely,   we provide the   total energy calculations. Moreover,  we will discuss the  energy minimization  needed  to predict structure stability behaviors.     In the case of   the double square geometries, we exploit the pseudo potential for the  germanium and  the silicon  with a group space associated with the  tetragonal type $P$ where the size parameters are considered as follows 
\begin{equation}
a=b=3,71.
\end{equation}
For  the double hexagonal structure, we use the pseudo potential  for the carbon and  the silicon  relaying   on a  type $P$ group space  with  the  hexagonal and the  trigonal  configurations. The used sizes are given by 
 \begin{equation}
a=b=4,26.
\end{equation} 
In both cases, the $c$ size  has been  considered with appropriate values to get 2D limit computations.  These  calculations  have been   performed,   with a plane-wave basis,  by using an energy cutoff of 500 eV.    To investigate the associated physical properties, we  use  K-mech  of  $8  \times 8 \times 1$.  Moreover, the present  computations  have been carried out in the plane
wave framework within the GGA approximations. To get rapid and well
estimated computations of such double structures, we first examine the total
energy in terms of the energy cut-off and K-points. It is denoted that 
the XCrySDen software has been exploited to visualize the positions of the
atom organizations and select the K-path to obtain such points. The self
consistent computations and the total energy calculations have been
performed for  a family of  A$k$B$k$  type structures where the atoms of type A belong to the small $k$-polygons and type B atoms are placed at the big ones. For
simplicity reasons, we consider the Si4Ge4 squares and the Si6C6 hexagons.
We anticipate that  the calculations could be done for other atom configurations
of  the A$k$B$k$ material structures exhibiting $D_{k}\times D_{k}$ dihedral
symmetries. In Fig.\ref{F7}, we present the cut-off energy and the K-mesh for
unit cells of the Si4Ge4 material structure via QEC computations. 
It has been remarked that the total energy decreases by increasing the cut
off energies until it gets stabilized at specific equilibrium
states. Similar behaviors have been observed for K points.

\begin{figure}[!ht]
\centering
\begin{tabular}{cc}
\includegraphics[scale=0.35]{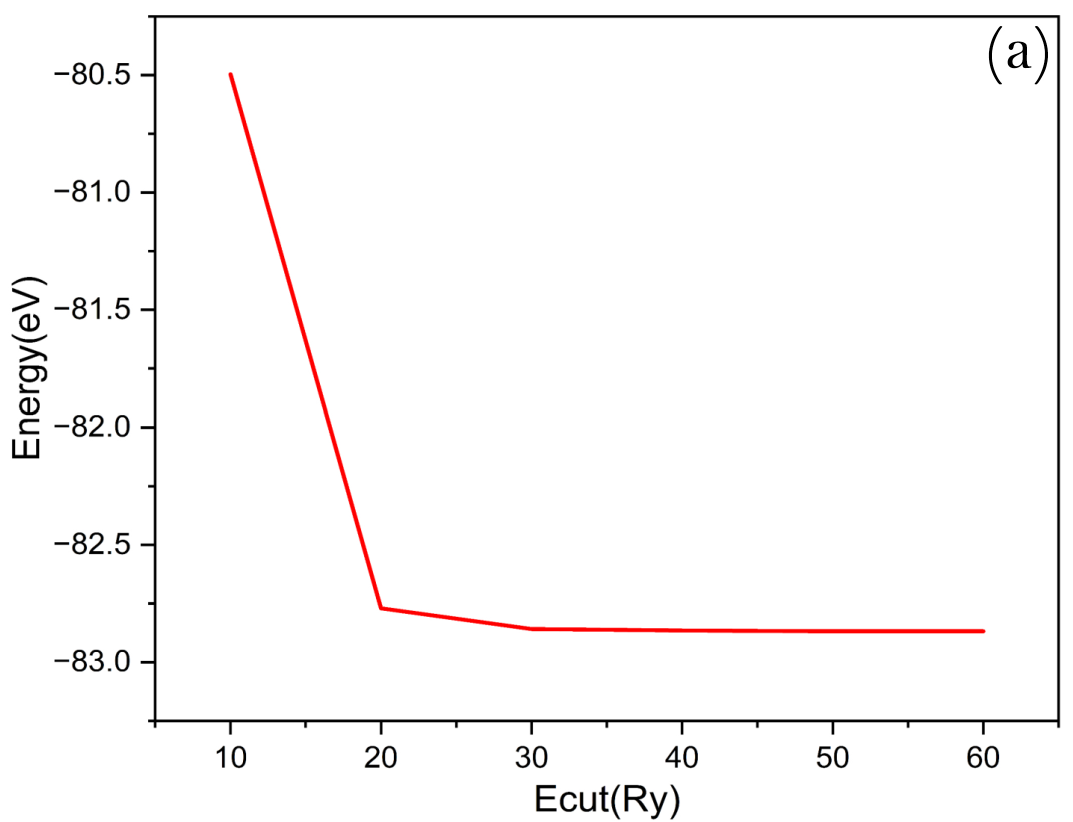} & %
\includegraphics[scale=0.35]{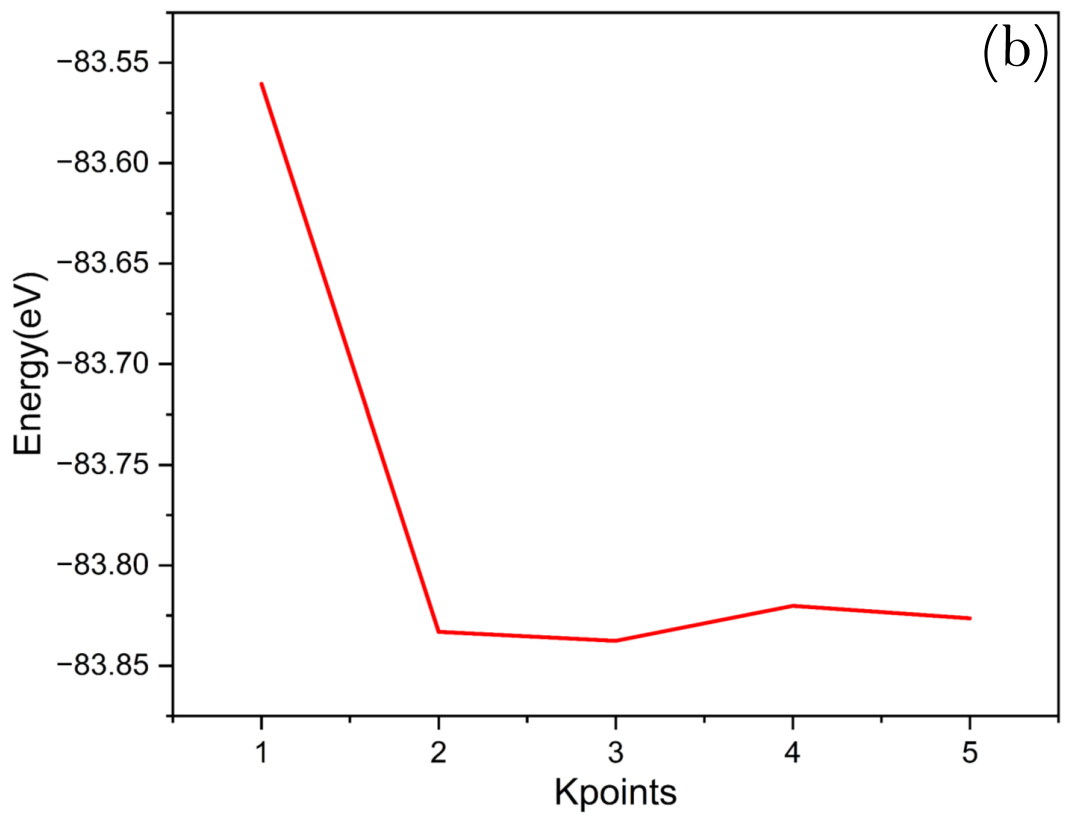}%
\end{tabular}%
\caption{(a): Total energy as a  function of cut off energy of  the Si6C6 material.
(b): Total energy as a function of K-points  of  the Si6C6 material.}
\label{F8}
\end{figure}

In Fig.\ref{F8}, we illustrate the cut-off energy and  the K-mesh  points for unit cells
of  the Si6C6  material by help of QEC computations.

A close examination shows that similar aspects have been obtained supporting
the next investigations on the associated physical properties.

\section{Results  and discussions of  opto-electronic properties  for   double structures}
Having discussed the stability behaviors of the proposed double structures,
we now  approach certain physical properties via QEC methods with the GGA
approximations. For simplicity reasons, we only consider the opto-electronic
properties by examining the electronic band structures, the total and the
partial densities of the states (DOS) and (PDOS). It has been understood
that such properties can be investigated by placing the studied materials in
the light source backgrounds. By considering linear homogeneous and
isotropic mediums, the relevant optical quantities can be computed via a
complex scalar $\epsilon(\omega)$ called the dielectric function depending
on the light frequency $\omega$. The associated calculations can be
performed using  the Kramers Kronig equations \cite{23,24,25}. This scalar
quantity, being decomposed as $\epsilon(\omega)=\epsilon_1(\omega)+i%
\epsilon_2(\omega)$, can carry information on the dispersion and the
electronic band structure behaviors. More details on the associated
computations could be found in many papers dealing with optical behaviors 
\cite{op}. To get more beautiful graphical representations, the optical
quantities will be given as functions of the energy.

\subsection{ DFT discussions  for 2D double square materials}

\begin{figure}[th]
\centering
\begin{tabular}{cc}
\centering \includegraphics[scale=0.35]{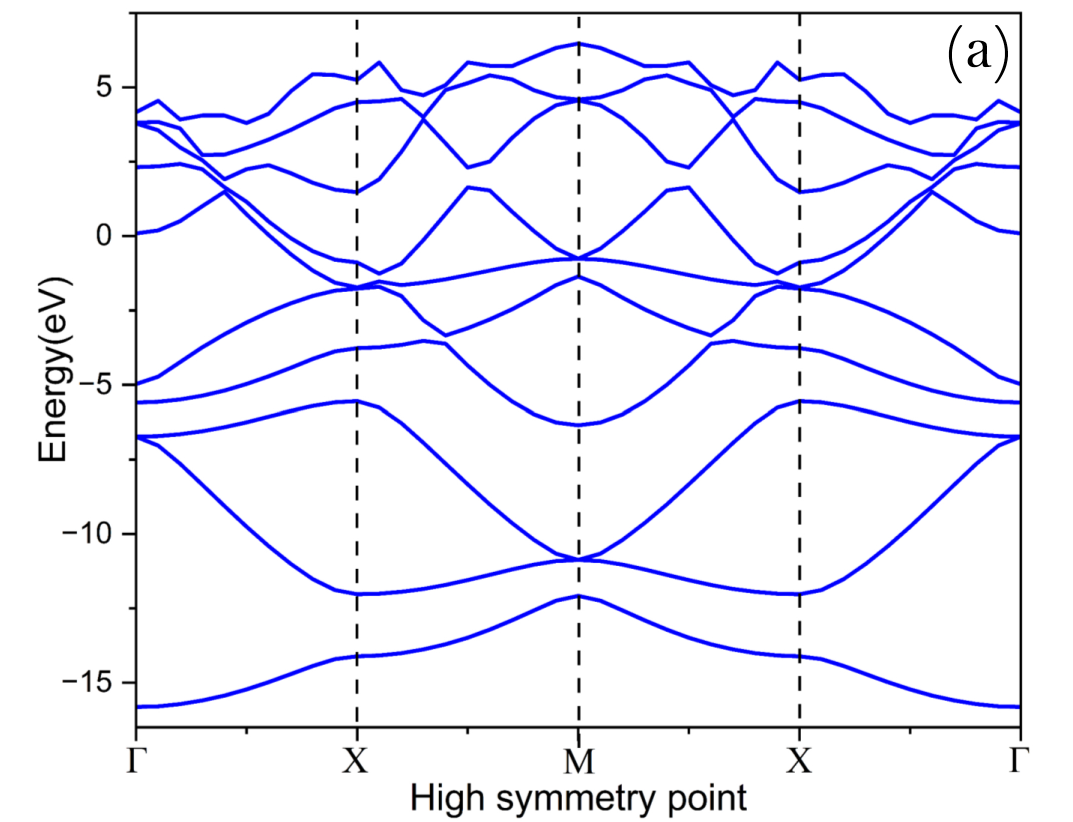} & %
\includegraphics[scale=0.35]{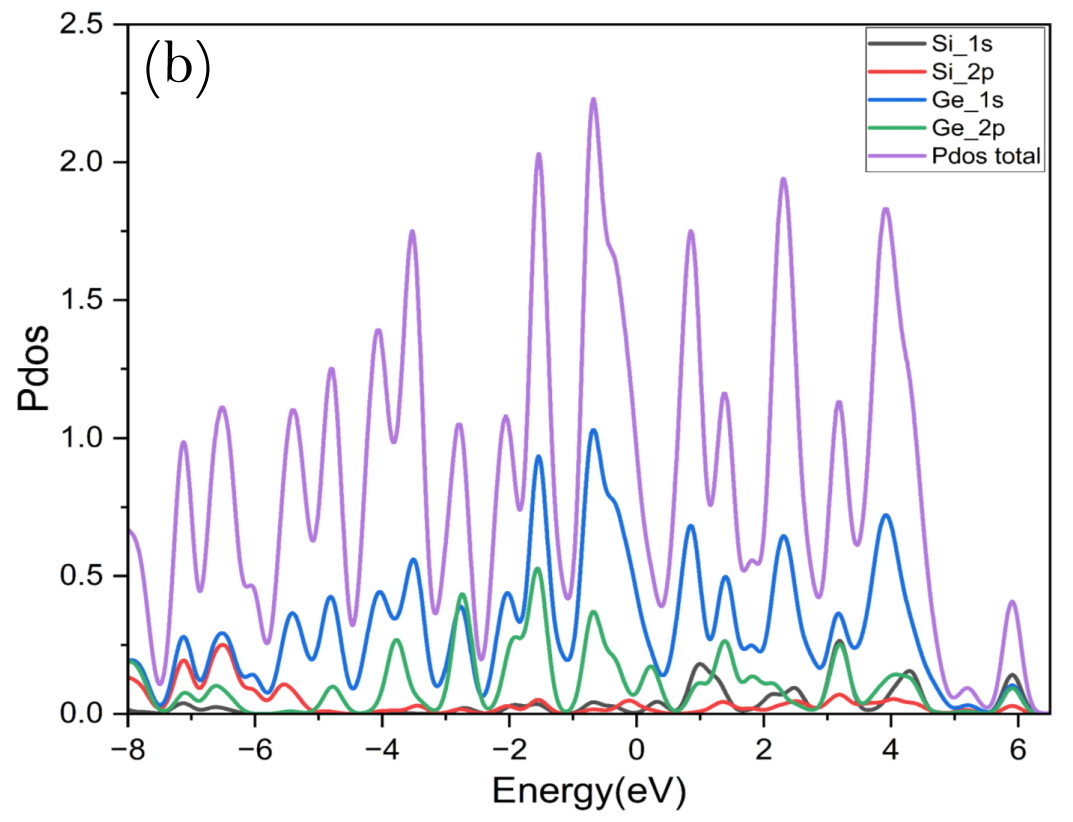}%
\end{tabular}%
\caption{(a): Band structure of Si4Ge4 material.  (b): Density of
states of Si4Ge4 material.}
\label{F9}
\end{figure}

It has been suggested that the determination of the electronic band
structure and the density of states is primordial for material
investigations. These studies are needed to reveal the potential
applications via the energy band gap calculation using the GGA
approximations. 
   Roughly,  we now provide  the band structure behaviors  of  the   Si4Ge4 material. Fig.(\ref{F9}) gives  its  electronic  band structure (a) and the partial  density 
states  (PDOS) (b).  
The figure \ref{F9} (a)   reveals clearly the overlapping between  the valence and  the conduction bands.  This means that    the present material 
exhibits  a metallic  behavior. The latter has been confirmed by   the  PDOS  computations  illustrated  in  the  figure \ref{F9}(b).  Similar behaviors have been obtained in a recent work dealing with a novel silicene-like material with a  pentahexoctite-silicon  geometry using DFT scenarios \cite{novel}.   This could  suggest  that  the   Si4Ge4 material  could  open gates for potential  applications in  the electronic activities.

The calculation of the dielectric function is important in order to show the
optoelectronic device applications. This quantity involves the physical
information on the response of the materials in the presence of the 
electromagnetic sources in terms of the interaction of the light rays and
the involved material atoms. Before analyzing the optical aspect, it is
important to mention that the variation could be performed either in terms
of the wavelength or the energy. Based on the fact that they are linked, the
energy will be a relevant variation.

\begin{figure}[th]
\centering
\begin{tabular}{cc}
\centering \includegraphics[scale=0.45]{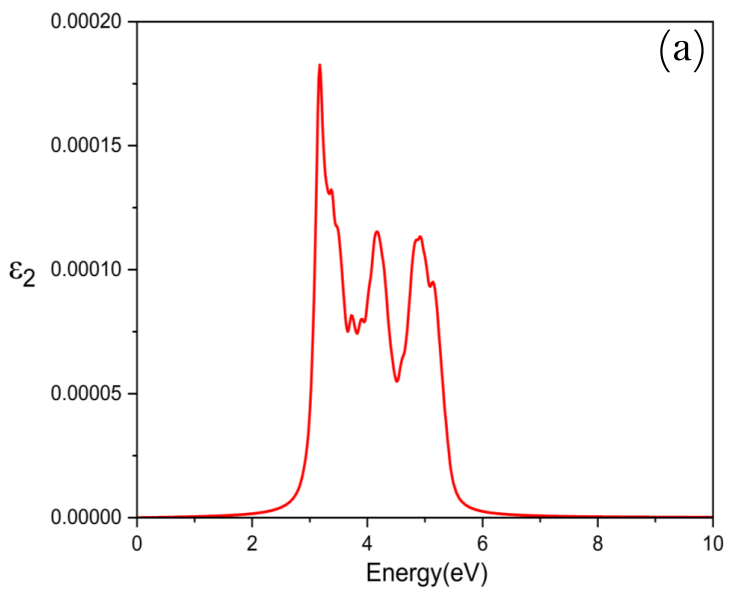} & %
\includegraphics[scale=0.45]{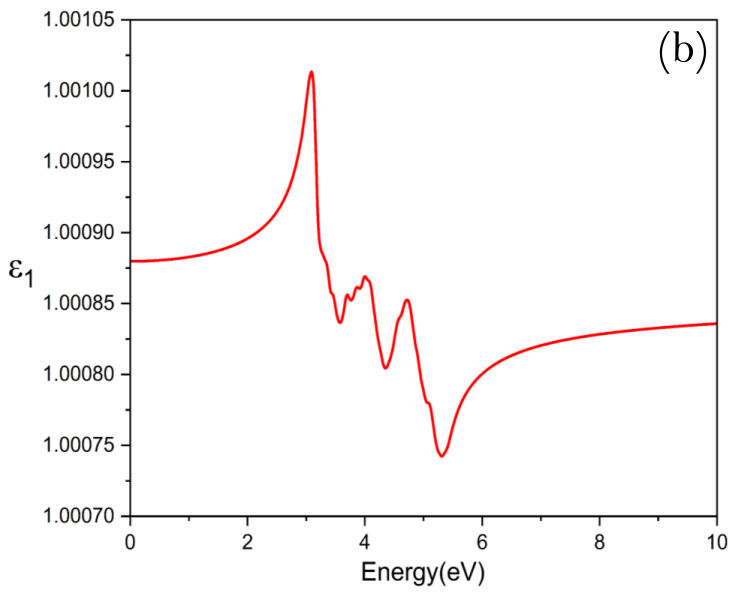}%
\end{tabular}%
\caption{(a):   Imaginary part of the
dielectric function in terms of the energy for the Si4Ge4 structure. (b):  Real part of the dielectric function  in terms of  of the
energy for the Si4Ge4 structure }
\label{F10}
\end{figure}

Fig.\ref{F10} illustrates the variation of the real and the imaginary
parts of the dielectric function as a function of the energy for the Si4Ge4
structure, respectively.
According to  the  figure \ref{F10}(b),  the static dielectric constant $\epsilon_1(0)$ value for the Si4Ge4 structure  is approximately 1  at the zero frequency limit.
It follows  also  that the real part $\epsilon _{1}(\omega )$
increases with energy. Then, a  first peak appears  around 3 eV followed by a  decreasing behavior.
The imaginary part of  the dielectric function is presented in the  figure \ref{F10}(a) showing an electromagnetic light attenuation in the modeled
material. It has been observed that the light absorption starts at a special
energy value.   The main peak of    $\epsilon _{1}(\omega )$  of   the Si4Ge4 material   is located   around   3 eV.   Moreover, it has been that the energies between    3 eV and   5,5 eV are relevant  for such a  optical quantity.

\begin{figure}[t]
\centering
\begin{tabular}{ccc}
\includegraphics[scale=0.278]{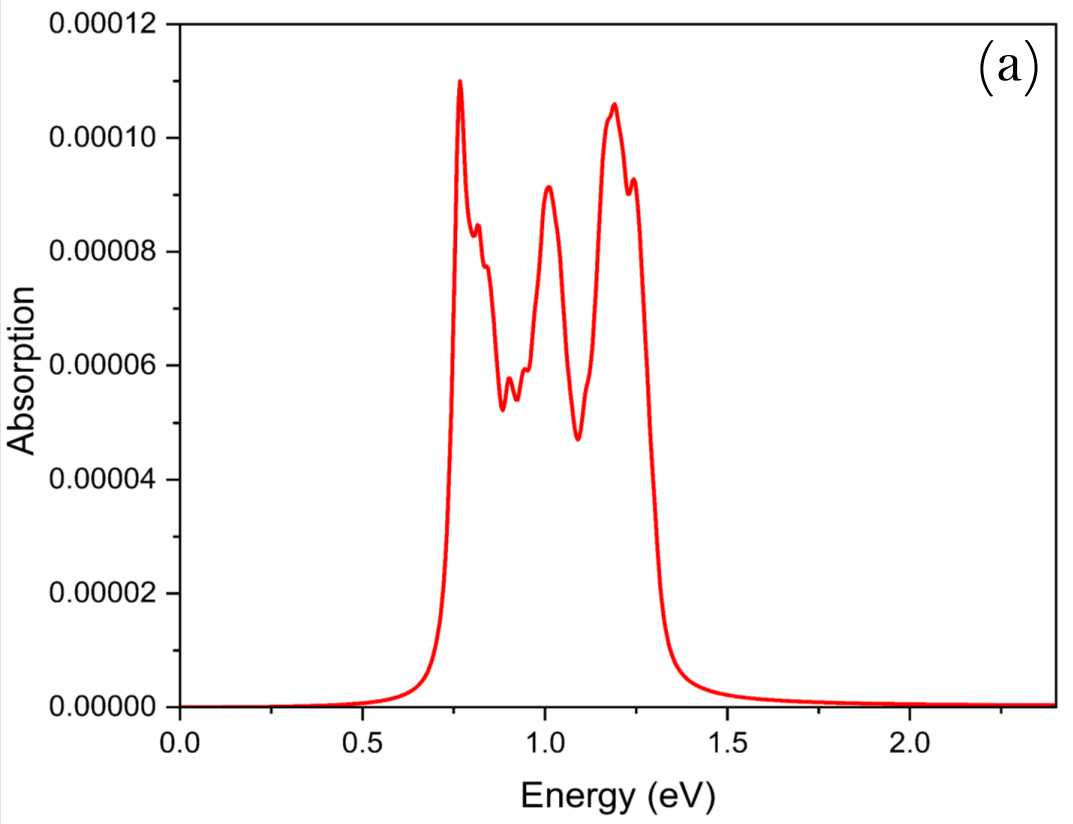} & %
\includegraphics[scale=0.40]{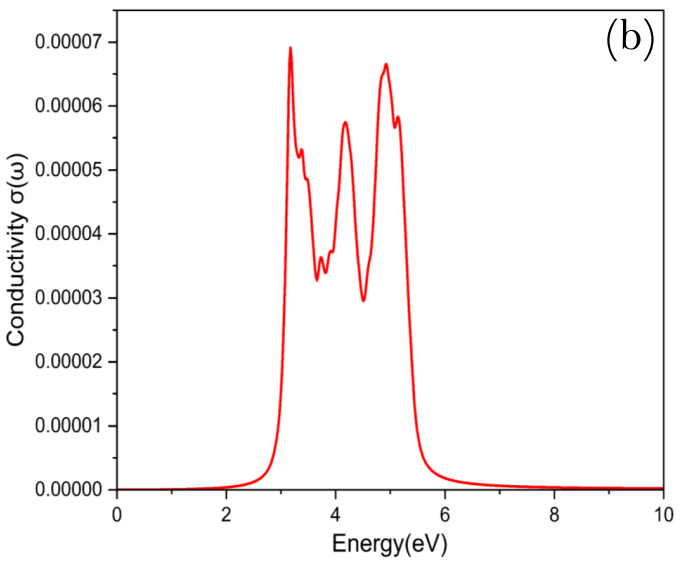} &  \\ 
\includegraphics[scale=0.40]{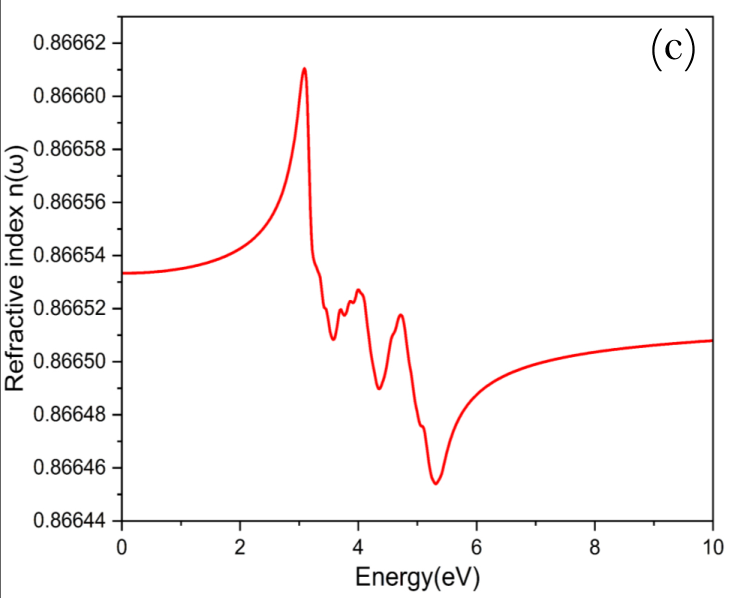} & %
\includegraphics[scale=0.40]{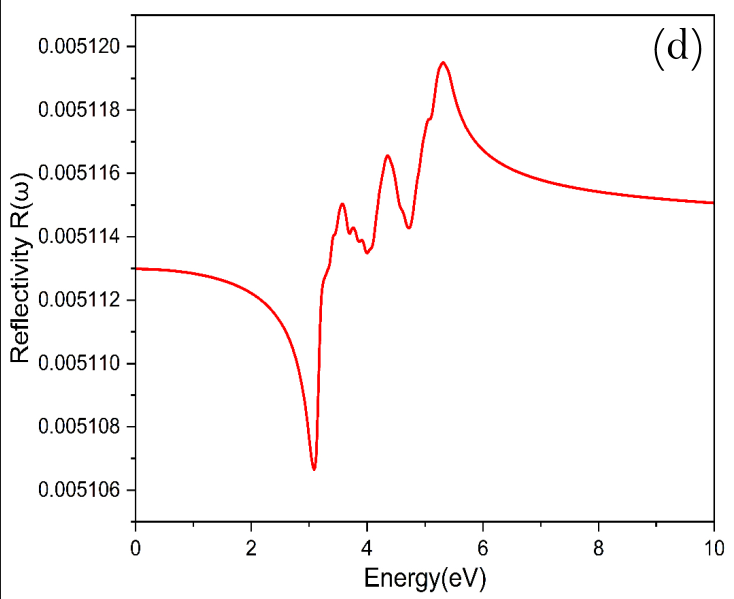} & 
\end{tabular}%
\caption{ (a): Absorption coefficient as a function of energy.   (b): Conductivity, as a
function of energy.  (c):    Refraction as a function of energy. (d): Refractive index as a function of energy. }
\label{F11}
\end{figure}

The calculations of the remaining optical quantities are crucial for the
optoelectronic device engineering applications. It has been remarked that
the photo-absorption coefficient, being obtained from the dielectric
function, is considered as the relevant one. Concretely, it enables a
material to capture light rays.  Fig.(\ref{F11}) illustrates the obtained results.

 Indeed, the figure \ref{F11}(a) shows the
variation in terms of the energy. It has been observed that this quantity
starts at a certain energy value being the optical gap. The 
figure \ref{F11}(b) illustrates the optical conductivity in terms of the energy.
Similarly as the absorption coefficient, the corresponding curve starts at
the optical band gap. The figure \ref{F11}(c)  represents the
refractive index $n(\omega )$ which characterizes the transparency degree  
of the material under the effect of the electromagnetic field. A rapid
examination shows that the associated curve shares similarities with the
real part of the dielectric function given in  the figure \ref{F10}(b). Moreover, it
takes a maximal value for lower energies. Concerning the reflectivity, it
has been displayed in the figure \ref{F11}(d). In the range
between 3 and 6 eV, it increases with energy.


\subsection{ DFT discussions for 2D double hexagonal materials}

\begin{figure}[!ht]
\centering
\begin{tabular}{cc}
\centering \includegraphics[scale=0.55]{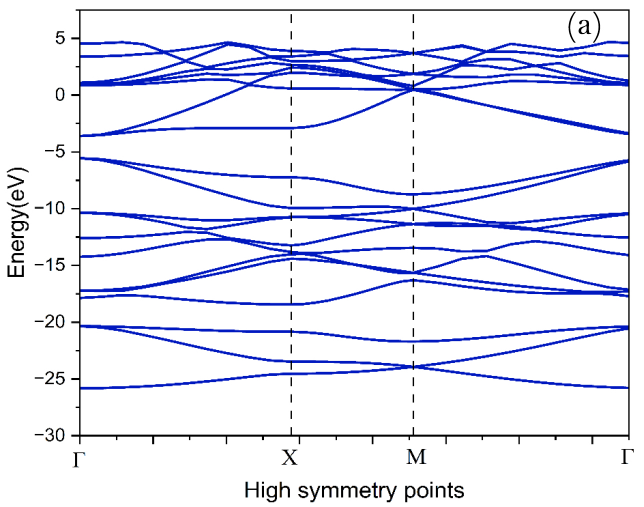} %
\includegraphics[scale=0.475]{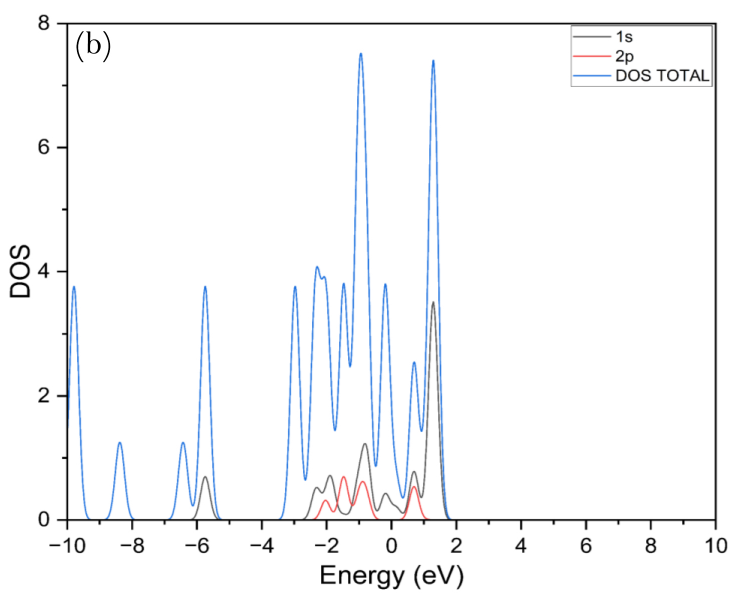} & 
\end{tabular}%
\caption{(a): Band structure for the Si6C6 material. (b):
Density of states for the Si6C6 material.}
\label{F12}
\end{figure}

Here, we consider the double hexagonal structure inspired by  the $G2$
Lie algebra. In particular, we provide numerical calculations for the Si6C6 material. It has been suggested that the determination of the electronic
band structure and the density of states is primordial for such invitations.
This study is needed to reveal the potential applications via the energy
band gap calculation using the GGA approximations. Fig.(\ref{F12}) gives the
band structure and the density of state for the Si6C6 material.

\begin{figure}[!ht]
\centering
\begin{tabular}{cc}
\centering \includegraphics[scale=0.45]{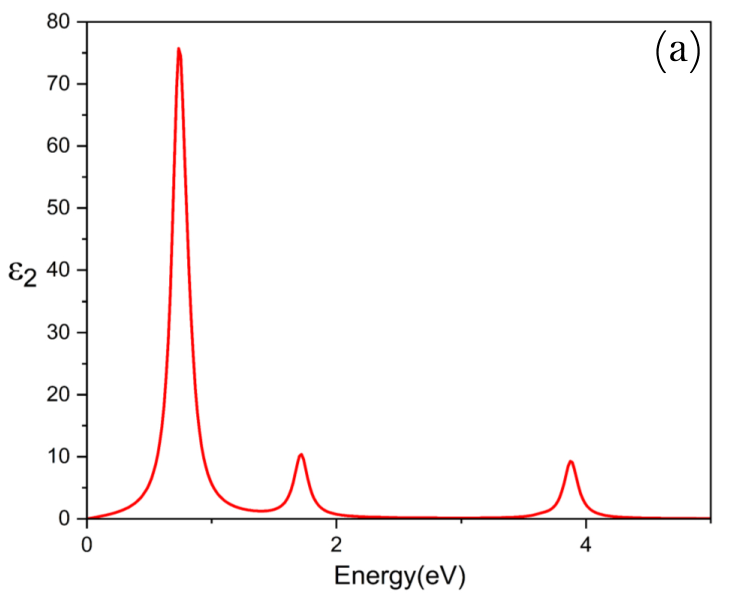} & %
\includegraphics[scale=0.45]{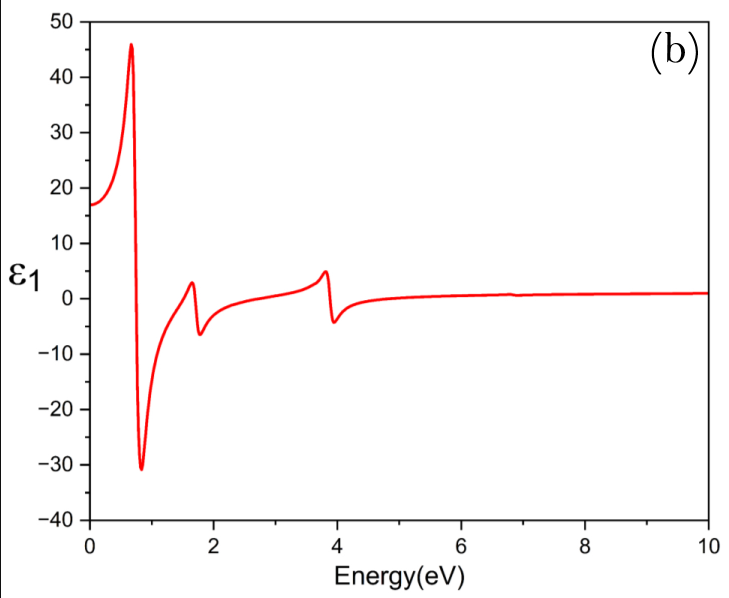}%
\end{tabular}%
\caption{(a): Real part of the dielectric function of  the Si6C6 compound  as a 
function of the energy. (b): Imaginary part of the dielectric
function of the Si6C6 compound as a function of the energy.}
\label{F13}
\end{figure}

 As before,  the  figure \ref{F12}(a)   shows clearly the overlapping between  the valence and the  conduction bands indicating that    the  Si6C6 material 
exhibits  a metallic  behavior. This property  has been  supported  by   the  PDOS  calculations    given in   the  figure \ref{F12}(b). This 
 metallic behaviors for such  a  double hexagonal
structure could be exploited for certain electronic applications. We move now to the optical behavior which should be approached
via the dielectric function being  represented in Fig.(\ref{F13}). 
This figure illustrates  the variation of  the  real and the imaginary parts as
a function of the energy.     Indeed, the figure \ref{F13}(b) shows that the static constant 
value $\epsilon_1(0)$   for  the Si6C6  material   is 16,5 eV.  For small energy values, 
 the spectrum of $\epsilon_1(\omega)$ increases. Then,   it decreases until it  gets 
 negative  values.  On the other hand, the figure \ref{F13}(a) illustrates  the 
$\epsilon_1(\omega)$  curve starting from o eV.  This  confirms the previous one obtained from   the band structure and  the PDOS  computations showing  expected behaviors.

\begin{figure}[t]
\centering
\begin{tabular}{ccc}
\includegraphics[scale=0.45]{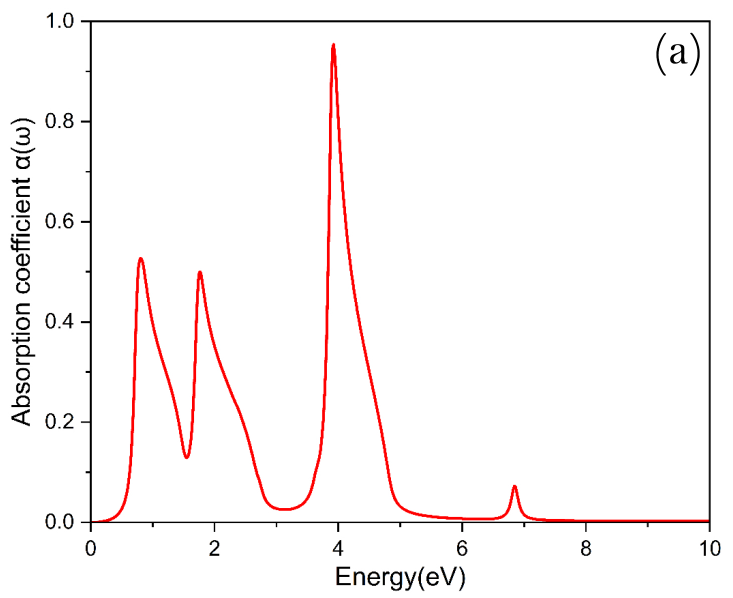} & %
\includegraphics[scale=0.45]{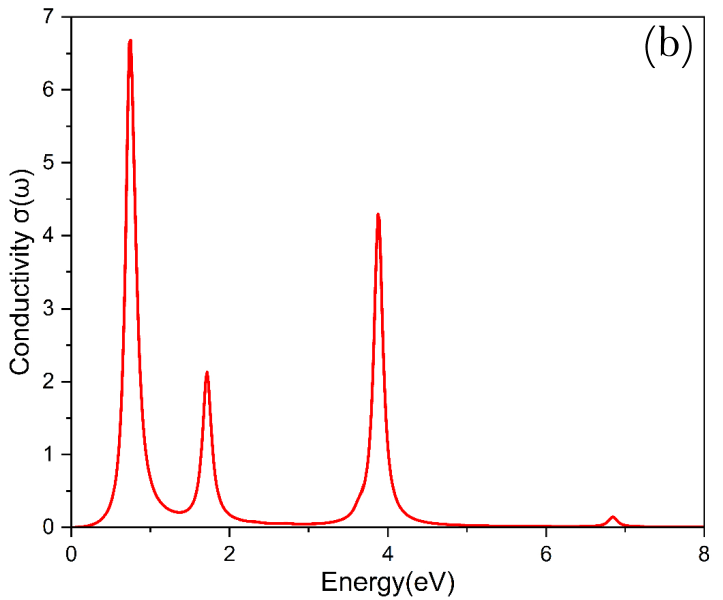} &  \\ 
\includegraphics[scale=0.45]{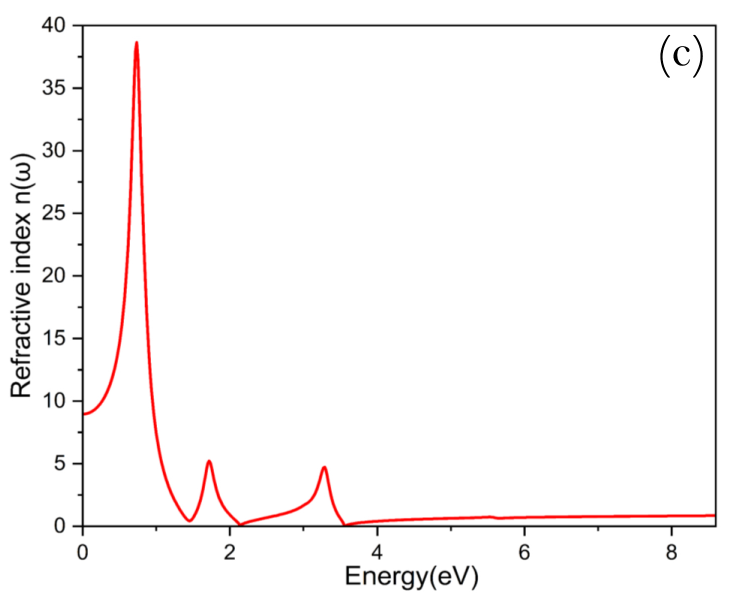} & %
\includegraphics[scale=0.45]{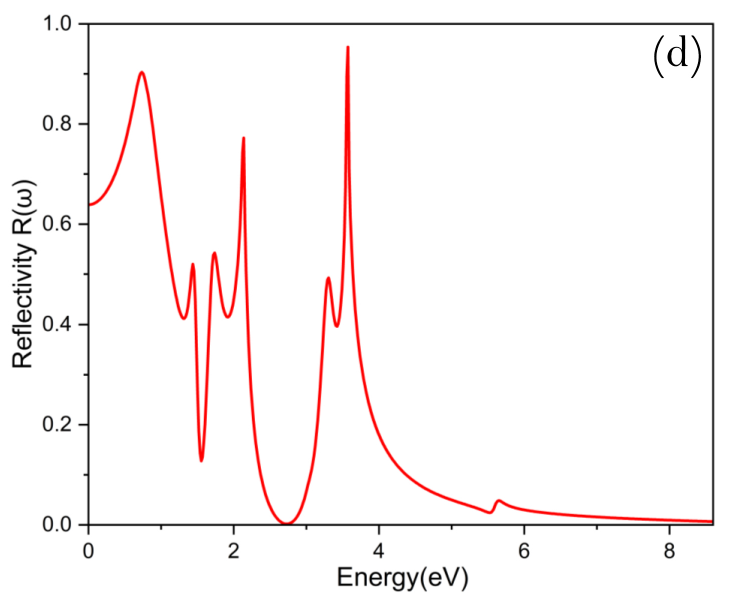} & 
\end{tabular}%
\caption{ (a): Absorption coefficient of Si6C6 as a function of energy.  (b):
Conductivity of Si6C6 as a function of energy. 
(c): Refraction of Si6C6 as a function of energy.  (d): Refractive
index of Si6C6 as a function of energy. }
\label{F14}
\end{figure}

Concerning the rest of the optical quantities, they are  depicted  in Fig.(%
\ref{F14}).

The graphs of this  figure confirm the electronic results showing metallic behaviors. For
certain applications, the obtained numerical findings could be refined by
considering other atomic configurations. This  needs  more serious
and deep thinking via collaborations with experts on the associated field.
\newpage
\section{Conclusion and concluding suggestions}

Motivated by the fact that the  graphene and the  silicene involve more attractive
physics, and inspired by the hexagonal   geometry of the hadronic structure in
the QCD sector of the SM described by the corresponding  symmetries as
well as the possible other underlying symmetries, we have  modeled  and simulated  2D 
materials inspired   by rank 2 non-simply laced Lie algebras. The associated
geometries are based on double periodic structures in the flat plane.
Concretely, we have investigated  two geometries relaying on 
squares and hexagons  exhibiting  the $D_{4}\times D_{4}$ and  the $D_{6}\times D_{6}$
dihedral group invariances, respectively. After rational studies of  2D material
building models, we have provided a numerical investigation  of  the  Si4Ge4 and  the Si6C6 compounds  using DFT
scenarios via QEC with GGA approximations. Inspired by the  $so(5)$ and $G2$ Lie
algebras, we have discussed the opto-electronic behaviors of 2D materials
with double squares and hexagons, respectively.    After stability discussions,  we have shown that such inspired Lie algebra 2D materials exhibit metallic  behaviors
which could be exploited for certain electronic applications.  Then, we have computed and discussed  the   relevant the optical quantities  such as 
the absorption spectra, the dielectric function, the refractive index,   and  the  reflectivity. These quantities  have been plotted and analyzed.    We have revealed  that the
obtained results could provide interesting results going beyond the single
structure associated with simply laced Lie algebras. The present
calculations could be extended in many directions. Precisely, they could be
generalized to other DFT numerical codes. Moreover, the present studied
physical properties could be also extended by considering other properties such as  thermo-mechanical
behaviors.      It has been suggested that  the  dynamic stability through phonon calculations could be performed. However, such computations  may need  powerful simulations. We leave  such an  investigation  for future works by means of   sophisticated  simulation machines. 
The obtained results  could supply certain roads for other investigation works supported by  appropriate applications.

This work comes up with extra remarks and questions. A natural question
concerns rank 3 Lie algebras which could provide data on 3D materials.
However, the classification of such Lie algebras is still an open question.
Motivated by the link between graphene and HEP, partial results could bring
interesting findings in lower dimensional materials. Supported by analogue
gravity scenarios, this could furnish a curiosity to go beyond standard
results. It is possible to propose certain duality approaches which could
relate  lower dimensional strongly coupled systems to gravity theory via BH
physics. Moreover, it has been speculated that quantum gravity in lower
dimensions could find place in lower dimensional materials. We anticipate that the computations
explored in the study of thermodynamic and optical behaviors of BH could be
exploited  in material physics by building bridges between such different
physical domains \cite{sss,ttt}.

Another relevant issue  could relay on the potential applications of the
double structure since 2D materials have been exploited in many
investigation {paths. We expect that the double structure could be exploited
to generate alternative  routes in finding certain applications associated with 2D
materials.      We could   anticipate  that  the double structure could be adapted to all two dimensional materials involving  a single flat geometry.   More precisely,  a promising investigation direction may concern  the study of  the inter-distance effect  on parallel layers  of such Lie algebra  inspired materials.  A possible situation is to propose  bilayer-like  systems  by considering a dynamic monolayer moving  on an orthogonal direction  using the so-called  buckled geometries in certain   specific arrangements.  This could open  band gap by simply varying the distance providing alternative potential applications including photovoltaic cell modelings.  We hope to  address such questions in future works. 

Finally, we could   implement generalized Cartan  matrices derived from 
Calabi-Yau space geometries to provide new root systems generating  other material
structures \cite{emi}. We believe that all these questions could be
addressed in future works.

\section*{Acknowledgments}

The authors would like to thank theirs families for support and patience.
Precisely,  they would like to thank their mothers: Fatima (AB) and
Fettouma (SEE). They would like to thank also the all First year (M1) master
students (MMR-2022-2023) for interesting questions on Lie algebras and related topics.
Especially, they are grateful to A. El Azizi and K. Loukhssami for computer
science helps.   The authors would like to   thank the anonymous referees for interesting comments and suggestions. They would like  to thank also  H. Labrim for discussions.
\begin{appendices}
\section{Lie Algebras}
 In this appendix, we give a concise review on Lie algebras by explaining the mathematical terms used  in the present work.  More details  could be found in  \cite{3,3c}.  It is recalled that such algebras play a primordial role in the understanding of the matter and radiation of the Universe via an elegant  framework called Standard Model. Moreover,  the theory of such algebras has been also explored in the condensed matter physics.  Roughly,  we provide certain useful definitions. 
 A Lie algebra  Lie $L$  is a vector space  on  the complex   or numbers  endowed by 
  a bracket $\left[ ,\right] $ 
 \begin{eqnarray*}
 \left[ ,\right]: \quad  L. L&\rightarrow & L\\
 (x,y) &\rightarrow& \left[ x,y\right]=xy-yx
\end{eqnarray*}   satisfying three conditions 
\\ 1.  bilinearity  
\begin{eqnarray*}
\left[ \alpha x+\beta y,z\right] =\alpha \left[ x,z\right] +\beta \left[ y,z%
\right]  \end{eqnarray*}
where   $\alpha$ and  $\beta $ are complex scalars. \\
 2. antisymmetry 
\begin{eqnarray*}
\left[ x,y\right] =-\left[ y,x\right]
\end{eqnarray*} 3.
Jacobi identity
\begin{eqnarray*}
\left[ \left[ x,y\right],z\right] +\left[ \left[ y,z\right],x\right] +%
\left[ \left[ z,x\right],y\right] =0 \end{eqnarray*}  for all $x$,
$y$, and $z$ inside  $L$. It is recalled that the dimension of a Lie algebra $L$ is the dimension as a vector space. 
A close examination shows that there are beautiful examples given  in terms of matrices called Lie algebras of matrices.  The most studied ones are $su(n)$, $so(2n)$  and $so(2n+1)$  Lie algebras. All these Lie algebras can be considered   as sub-algebras of  the $g\ell(n)$  Lie algebra  being  the  $n \times n$ matrix algebra with  the matrix multiplication  of dimension $n^{2}$.  A generic element  of such a Lie algebra reads as 
\begin{eqnarray}
X=(x_{ij}), \qquad i,j=1,\ldots, n
\end{eqnarray}
without any constraint.
\subsection{ $su(n)$ Lie Algebras}
These  Lie algebras are defined in terms   of  the $n \times n$  traceless matrices with  the anti-hermitian   condition
\begin{eqnarray*}
 su(n)=\left\{ X \in g\ell(n),\;\;
X+X^{+}=0,\:\;trX=0\right\}
\end{eqnarray*}
where $+$ is the adjoint action. It has been shown that  the dimension of the  $su(n)$  Lie algebra is $n^{2}-1$.  The leading example of such a family  is  the $su(2)$ Lie algebra   defined by
\begin{eqnarray*}
 su(2)=\left\{ X \in g\ell(2), \;\;X+X^{+}=0 ,\;\; trX=0\right\}
\end{eqnarray*}
which is a 3  dimensional  Lie algebra.  The latter has been explored in many places including quantum mechanics and particle physics.  In particular, it has been used to describe  spin particle representations. Another beautiful one  is  the  $su(3)$ Lie algebra defined by 
\begin{eqnarray*}
 su(3)=\left\{ X \in g\ell(3), \;\;X+X^{+}=0 ,\;\; trX=0\right\}
\end{eqnarray*}
involving a  hexagonal root  system.
\subsection{$so(2n+1)$ Lie Algebras}
To define such Lie algebras,  one needs  a specific matrix   $J$  taking the following form
 \begin{eqnarray*}
 J=\left(
\begin{array}{ccc}
1 & 0& 0 \\
0&0_{n\times n} & I_{n\times n} \\
0& I_{n\times n} & 0_{n\times n}
\end{array}%
\right)_{2n+1\times 2n+1}.
 \end{eqnarray*}
In this way, one can define  the   $so(2n+1)$ Lie algebras   as 
 \begin{eqnarray*}
 so(2n+1)=\left\{ X\in g\ell(2n+1),\quad JX+X^{t}J=0_{2n+1\times 2n+1}, \;\;\; Tr X=0 \right \}
 \end{eqnarray*}
where $t$  denotes  the orthogonal  condition. It is denoted that $n=2$ provides  the so(5) Lie algebra used in the section 2.  It is worth noting, in passing,  that one can define  the  $so(2n)$  Lie Algebras via the matrix 
\begin{eqnarray*}
 J=\left(
\begin{array}{cc}
0 & I_{n\times n} \\
I_{n\times n} & 0%
\end{array}%
\right) _{2n\times 2n}.
\end{eqnarray*}
Precisely,  they  are  defined as follows 
\begin{eqnarray*}
so(2n)=\left\{ X\in g\ell(2n), \;\; J X+X^{t}J=0_{2n\times 2n},\;\;\; Tr X=0\right\}.
\end{eqnarray*}  
A special case is associated with     $n=4$ producing the  $so(8)$ Lie  algebra.   To make contact with physics, one usually uses the root system and the representation  concepts.  It is recalled that the root system  encodes many data on the associated Lie algebras  which could be exploited in many physical applications. Certain techniques  dealing with  roots   could generate new Lie algebras. The fundamental example is the $G2$ Lie algebra which can be derived from  the  $so(8)$ Lie  algebra using the folding mechanism in the graph theory  representation known by Dynkin diagrams. Indeed,  it has been shown that the Dynkin diagram  of the  $so(8)$ Lie  algebra involves a  trivalent geometry.   Identifying   the  permuted   external nodes  by the symmetric group $S_3$, one  generates the  $G2$ Lie algebra  graph encoding the most important features.  More details on such a scenario  can be found in  \cite{3}. 
\end{appendices}

\end{document}